# Mid-infrared Chemical Nano-imaging for Intra-cellular Drug Localisation.

William S. Hart[1,*], Hemmel Amrania[1], Alice Beckley[2], Jochen R. Brandt[3], Sandeep Sundriyal[3,4], Ainoa Rueda-Zubiaurre[3], Alexandra E. Porter[5], Eric O. Aboagye[2], Matthew J. Fuchter[3], and Chris C. Phillips[1,*]

[1] *Experimental Solid State Group, Department of Physics, Imperial College London, London, SW7 2AZ, UK*

[2] *Department of Surgery and Cancer, Imperial College London, London, UK*

[3] *Department of Chemistry, Imperial College London, London, SW7 2AZ, UK*

[4] *Department of Pharmacy, Birla Institute of Technology and Science Pilani, Pilani Campus, 333031, Rajasthan, India.*

[5] *Department of Bioengineering, Royal School of Mines, Imperial College London, London, SW7 2AZ, UK*

[*] Authors to whom correspondence should be addressed. Electronic mail:

chris.phillips@imperial.ac.uk

**Summary Paragraph.**

**In the past two decades a range of fluorescence cell microscopy techniques[1] have been developed which can achieve ~10nm spatial resolution[2,3], i.e. substantially beating the usual limits set by optical diffraction. However, these methods rely on specialised labelling[4]. This limits the applicability, risks perturbing the biology [5,6] and it also makes them so-called "discovery techniques" that can only be used when there is prior knowledge about the biological problem. The alternative, electron microscopy (EM), requires complex and time-consuming sample preparation, that risks compromising the sample's integrity. Samples have to withstand vacuum, and staining with heavy metals to make them conductive, and give usable electron-contrast. None of these techniques can**

directly map out drug distributions at a sub-cellular level. Recently infrared light-based scanning probe techniques have demonstrated a capability for ~1nm spatial resolution [7,8]. However, they need samples that are flat, dry and dimensionally stable and they only probe down to a depth commensurate with the spatial resolution, so they yield essentially surface chemical information. Thus far they have been applied only to artificially produced test samples, e.g. gold particles[9], or isolated proteins[10] on silicon. Here we show how these probe-based techniques can be adapted for use with routinely prepared[11] general biological specimens. This allows for "Mid-infrared Chemical Nano-imaging" (MICHNI) that delivers chemical analysis at a ~10 nm spatial resolution, suitable for studying cellular ultrastructure. We demonstrate its utility by performing label-free mapping of the anti-cancer drug Bortezomib (BTZ) within a single human myeloma cell. We believe that this MICHNI technique has the potential to become a widely applicable adjunct to EM across the bio-sciences.

Our MICHNI approach is based on so-called "scattering-type scanning near-field optical microscopy[12] (s-SNOM). An atomic force microscope (AFM) is used to position a sharp conducting probe to within a few nm of the object to be imaged, and a laser is focussed on the region where the two meet (Fig.1). The sharp probe tip concentrates the optical field, by a nanoscale analogue of the "lightning rod" effect, to a region commensurate with its radius. An analysis of the backscattered light allows the optical response of the material just under the tip[12] to be measured. This gives an optical image whose spatial resolution is determined by the tip radius, and not the light wavelength[13]. A spatial resolution of <10nm is routinely available with robust, commercially available probes. The optical image is built up by rastering the tip across the sample, and a standard AFM topography image is generated at the same time.

Previous s-SNOM studies have been restricted to narrow wavelength range studies of solid state systems. Studies include PMMA test samples[14,15], Quantum Metamaterial Superlenses[16] and oxide[17] and graphene[18] layers. In the biological realm, demonstrator trials have imaged cylindrical tobacco mosaic viruses deposited on Si [19], and dried proteins in lipid bilayer fragments extracted from bacteria and mounted on gold[20].

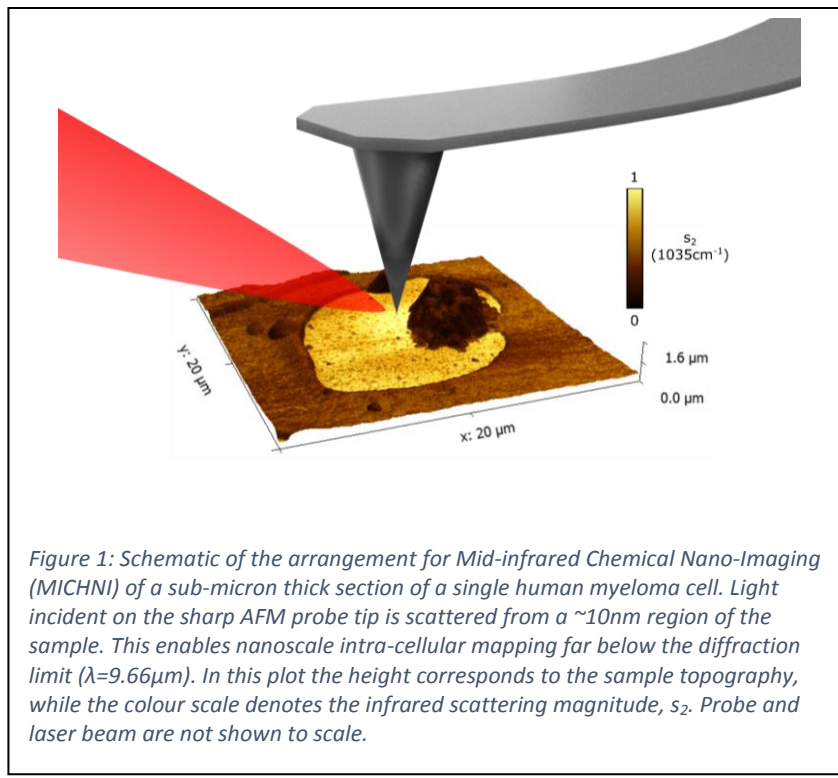

Figure 1: Schematic of the arrangement for Mid-infrared Chemical Nano-Imaging (MICHNI) of a sub-micron thick section of a single human myeloma cell. Light incident on the sharp AFM probe tip is scattered from a ~10nm region of the sample. This enables nanoscale intra-cellular mapping far below the diffraction limit ($\lambda$=9.66µm). In this plot the height corresponds to the sample topography, while the colour scale denotes the infrared scattering magnitude, $s_2$. Probe and laser beam are not shown to scale.

Mid-IR radiation in the so-called "chemical fingerprint" spectral region (approximately 5 µm < $\lambda$ < 12µm) is absorbed in localised bond-specific vibrational transitions that are characteristic of specific chemical moieties. The resulting absorption peaks are well-characterised, and have long been used, e.g. in Fourier transform infrared (FTIR) spectroscopy to identify and characterise (un-labelled) chemicals and to analyse mixtures[21] in chemistry and the life sciences.

If these same wavelengths are used for s-SNOM, the optical response of the region of the sample in the close vicinity of the tip is determined by these same bond-specific absorptions. This encodes the chemical mapping information into s-SNOM images[12].

Small-molecule drugs are exogenous species, in many cases containing chemical structures and functionality that give them a distinctive mid-IR "fingerprint" spectrum. This can make them stand out from the normal absorption signature of biological material, and makes them particularly suitable for mid-IR mapping (see supplementary information).

Here we show how a variant of the standard histopathological "formalin-fixed, paraffin-embedded" (FFPE) sample preparation method can be used to generate high-quality s-SNOM images from general biological material. We also show that, by adding a bank of widely tunable mid-infrared quantum cascade lasers (QCLs) that span the 5 μm < λ < 11μm "fingerprint" range[22,23], we can perform "Mid-infrared Chemical Nano-imaging" (MICHNI) in a way that allows us both to image chemical variations, and identify drug locations, in the cellular ultrastructure.

As a test example, here we have chosen to study the anti-cancer drug Bortezomib (BTZ). BTZ is a clinically approved agent for the treatment of multiple myeloma[24], and its mechanism of action proceeds through inhibition of proteasomes: cellular complexes that break down proteins. In addition to its clear therapeutic relevance, BTZ is particularly suitable for these experiments because it contains a B-C moiety which gives it a narrow vibronic absorption feature at a wavelength (1020 cm$^{-1}$) where there is little structure in the absorption spectrum of normal human tissue (see supplementary information).

Fig. 2 shows the AFM topography, and the second harmonic near-field scattered amplitude, $s_2$, of a thin (~1μm) section of a single RPMI-8226 myeloma cell, pre-treated with BTZ, fixed, dried and mounted on a glass microscope slide[12] (see methods section).

The AFM image (Fig. 2(a)) is essentially a height map, showing the profile of the biological material as it is distributed across the glass slide. It is generated mechanically, as the probe tip scans across the sample, like a nanoscale version of a record needle. Its lateral resolution is determined by the AFM tip diameter, and there is no chemical contrast. In comparison, the MICHNI image (Fig. 2 (b)) resolves a host of extra sub-cellular so-called "ultrastructure" features. These include details of the cell nucleus, various organelles and the membrane bilayer that have not been imaged before optically.

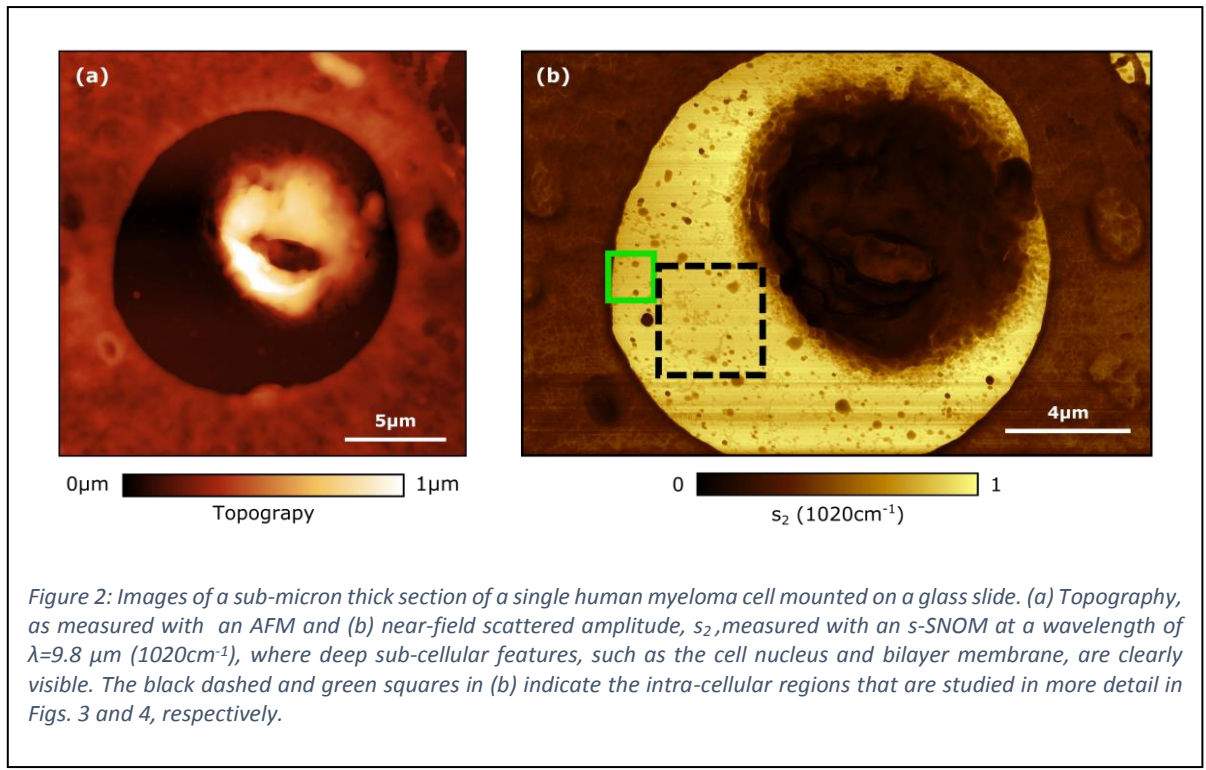

Figure 2: Images of a sub-micron thick section of a single human myeloma cell mounted on a glass slide. (a) Topography, as measured with an AFM and (b) near-field scattered amplitude, $s_2$, measured with an s-SNOM at a wavelength of $\lambda=9.8$ µm (1020cm$^{-1}$), where deep sub-cellular features, such as the cell nucleus and bilayer membrane, are clearly visible. The black dashed and green squares in (b) indicate the intra-cellular regions that are studied in more detail in Figs. 3 and 4, respectively.

Zooming in (Fig.3 (a) and (b)) on the black dashed square of Fig. 2(b) gives a visual indication of the resolution enhancement of the MICHNI image (Fig.3 (b)) of over the AFM one (Fig.3 (a)). Selecting a fine biological feature from the ultrastructure in the MICHNI image (Fig. 3(d)) and performing a line scan analysis shows how the "lightning rod" effect that concentrates the laser field beneath the tip in the MICHNI case has resulted in a ~3x resolution improvement over AFM. The ~10 nm MICHNI resolution, corresponds to $\sim \lambda/1000$ at the $\lambda = 9.8$µm imaging wavelength, i.e. beating diffraction by ~3 decades, and is comparable with current EM performance.

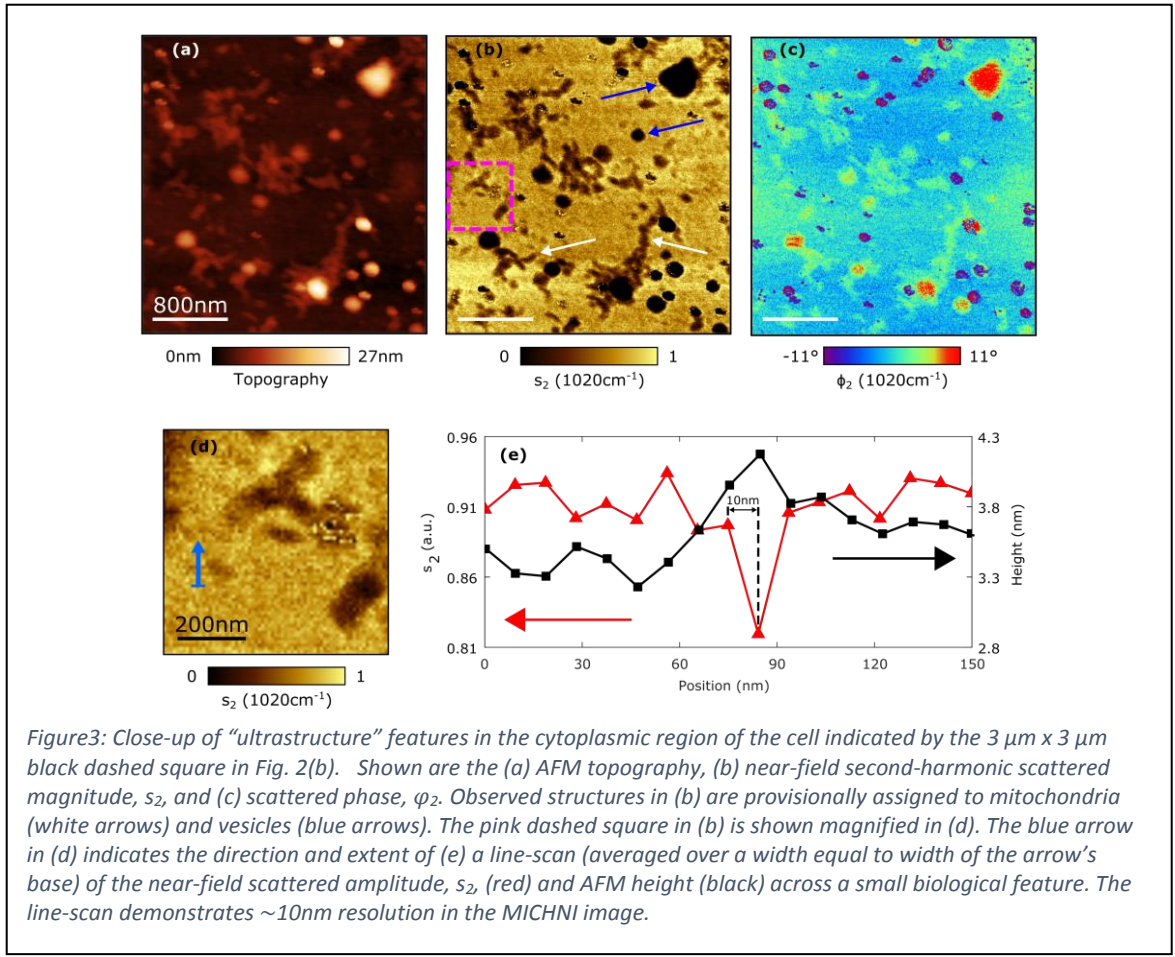

Figure3: Close-up of "ultrastructure" features in the cytoplasmic region of the cell indicated by the 3 µm x 3 µm black dashed square in Fig. 2(b). Shown are the (a) AFM topography, (b) near-field second-harmonic scattered magnitude, $s_2$, and (c) scattered phase, $\varphi_2$. Observed structures in (b) are provisionally assigned to mitochondria (white arrows) and vesicles (blue arrows). The pink dashed square in (b) is shown magnified in (d). The blue arrow in (d) indicates the direction and extent of (e) a line-scan (averaged over a width equal to width of the arrow's base) of the near-field scattered amplitude, $s_2$, (red) and AFM height (black) across a small biological feature. The line-scan demonstrates ~10nm resolution in the MICHNI image.

At this stage, identifying the various ultrastructure elements is complicated by the unknown amount of organelle shrinkage that is likely to occur during the drying stage of the sample preparation, but we tentatively identify oval mithochondria-like structures (white arrows, Fig. 3 (b)) which at ~ 250 nm long, are roughly 50% smaller than those seen in EM images of these cell lines[25]. Also present are a range of spherical vesicle-like structures with diameters in the 140nm-450nm range (blue arrows, Fig. 3(b)).

Looking ahead, we believe that specialised tissue processing protocols[26] will allow the morphology and scale of these structures to be better preserved during sample preparation, and that this, coupled with increased practitioner experience, will enable these sub-cellular features to be visually identified as confidently as with EM.

The frequency dependent phase of the backscattered light, $\varphi_2(\nu)$, is the key to the local chemical information that is contained in the optical near-field (see supplementary information). In order to isolate the BTZ contribution to the optical response, we first take a zero phase reference reading corresponding to a part of the image known to be glass (i.e. the lightest toned parts of the $s_2$-based images). This datum is used to generate a false-colour $\varphi_2(\nu)$ image (Fig.3c) where the regions of increased optical absorption due to BTZ cause $\varphi_2(\nu)$ to increase, and show up as red "hot spots" in the rendered image (see methods section).

For quantitative spectroscopic analysis, we zoom in further, to an area (Fig. 4) that corresponds the 1.4 µm x 1.4 µm green square in Fig. 2(b). The 125 nm x 125 nm cyan square and the 250 nm x 500nm white dashed rectangle show the acquisition regions for the near-field signal for BTZ and glass substrate respectively. Plotting the phase variation $\varphi_2(\nu)$, with the laser tuned to the absorption peak of the BTZ ($\nu = 1020$ cm$^{-1}$) revealed strong concentrations of the drug in some of the vesicle-like structures that were visible in the AFM image. This $\varphi_2(\nu)$ phase signal almost completely vanished when the laser was only slightly detuned (from $\nu = 1020$ cm$^{-1}$ to $\nu = 1000$ cm$^{-1}$) away from the BTZ absorption peak (Fig.4(c)).

An analytical treatment of the mechanism by which the tip backscatters light argues that the imaginary component of the scattered amplitude, Im($\sigma(\nu)$), (computed as described in the methods section) is proportional to the absorption coefficient of the sample in the tip region[12]. Fig 4(d) shows how Im($\sigma(\nu)$), averaged over the 125 nm x 125 nm cyan square, varies as the QCL lasers are tuned through the BTZ absorption peak. The optical response signal is averaged over a ~10 nm thick patch of material, and corresponds to a material volume of ~10$^{-22}$ m$^3$, (~0.1 fL).

The Im(σ(v)) spectrum in Fig. 4(d) (red circles) shows good qualitative agreement with the far-field absorption spectrum, (black line), which was obtained with a reference sample of pure BTZ measured in a conventional FTIR spectrometer (see methods section) Both spectra show a sharp peak at 1020cm$^{-1}$, with a difference in linewidth that is likely due to over-absorption in the reference sample whose optical thickness was not known due to the geometry of the "attenuated total reflection" method used.

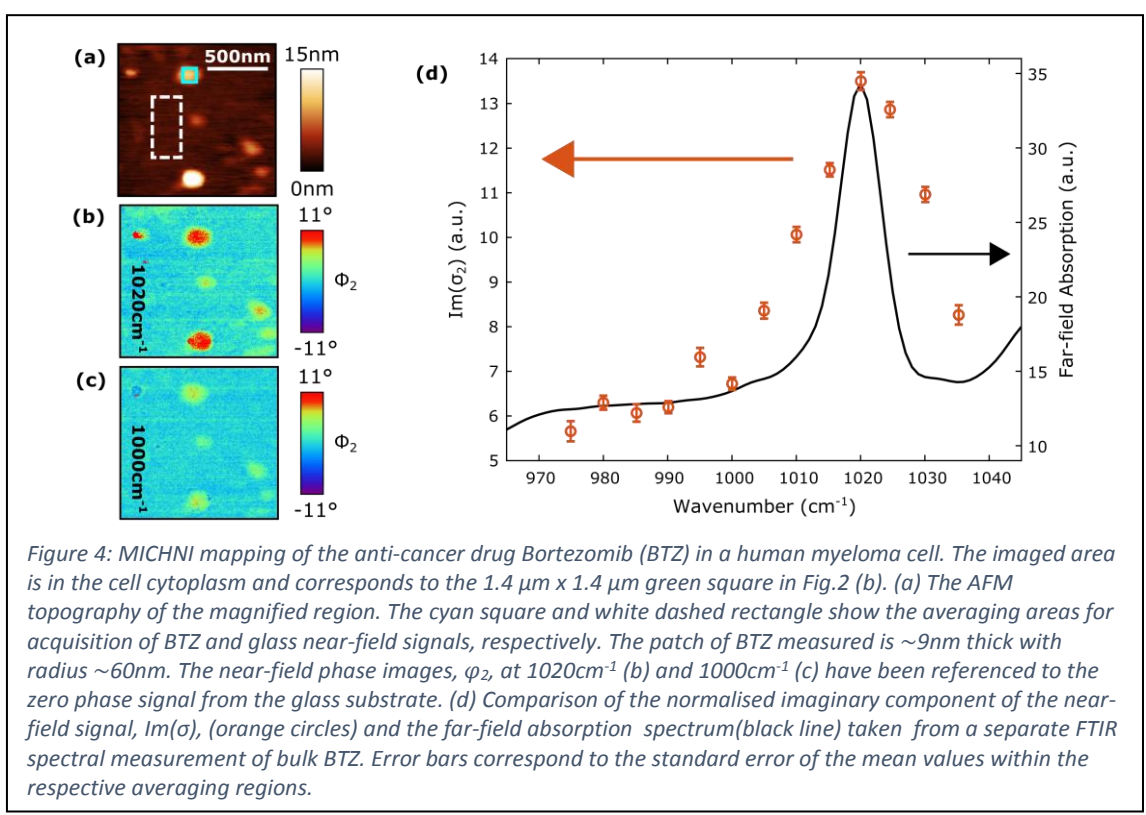

Figure 4: MICHNI mapping of the anti-cancer drug Bortezomib (BTZ) in a human myeloma cell. The imaged area is in the cell cytoplasm and corresponds to the 1.4 μm x 1.4 μm green square in Fig.2 (b). (a) The AFM topography of the magnified region. The cyan square and white dashed rectangle show the averaging areas for acquisition of BTZ and glass near-field signals, respectively. The patch of BTZ measured is ~9nm thick with radius ~60nm. The near-field phase images, $\varphi_2$, at 1020cm$^{-1}$ (b) and 1000cm$^{-1}$ (c) have been referenced to the zero phase signal from the glass substrate. (d) Comparison of the normalised imaginary component of the near-field signal, Im(σ), (orange circles) and the far-field absorption spectrum(black line) taken from a separate FTIR spectral measurement of bulk BTZ. Error bars correspond to the standard error of the mean values within the respective averaging regions.

As a proteasome inhibitor, one might *a priori* expect the BTZ signal to peak in regions of high proteasome concentration in the cell. Based largely on cell-fractionation studies[27], regions of proteasome activity and distribution are thought to typically include the outer endoplasmic reticulum, cytoskeleton and centrosomes, nuclei (but not nucleoli) and cytoplasm. We speculate that future MINCHI imaging studies will prove valuable in correlating proteasome localisation and inhibition in different cell types and under different conditions, thereby informing on the sensitivity of proteasome inhibition in certain disease states (i.e. multiple myeloma).

In conclusion, we have demonstrated MICHNI imaging of single cells for the first time, with a spatial resolution that already rivals EM and can probably be extended down to the ~1nm level with sharper tips[7,8]. The sample preparation protocol is simpler than for EM, and it preserves the sample chemistry and structure. The contrast mechanism is sufficiently well understood that quantitative information can be extracted from the MICHNI images and we have used this to map the distribution and localise the binding sites of an important drug within a single cell for the first time.

These findings promise a new route to studying intra-cellular chemical processes in a way that does not unintentionally influence them with artificial labelling. We believe that the MICHNI technique on its own has the potential for widespread applications across the fields of cancer, cellular pathology in general, and drug discovery. In addition, it could offer useful correlative morphological and chemical information for research problems that are currently being studied by EM.

**Methods**

**Spectroscopic Mid-Infrared CHemical Nano-Imaging (MICHNI).**

The MICHNI images presented in this study were generated using ~100mW level tuneable (~900-1900$cm^{-1}$) mid-infrared laser light generated by a switchable bank of commercial quantum cascade lasers (MIRcat, Daylight Solutions). This beam was focussed to the tip of a Pt/Ir-coated Silicon AFM probe (Arrow-NCPt, Nanoworld) in a commercial s-SNOM (NeaSNOM, Neaspec). The probe tip has a radius ~10nm, which acts to greatly enhance the scattering of the optical near-field in the region immediately beneath its apex[12]. The AFM component of the system was set to give a mean probe height of ~50 nm above the sample, and it was oscillating vertically with an amplitude of ~50nm at its mechanical resonance frequency $\omega \approx 280$ kHz.

The IR light backscattered from the tip was detected by a LN$_2$ cooled HgCdTe detector. The amount of scattered near-field optical signal depends on the tip-sample separation in a highly non-linear fashion, and this generates components at harmonics ($n\omega$) of the tip vibration frequency in the power spectrum of the backscattered light. Analysis of the "approach curves" [13] implies that, in general the higher the harmonics contain a larger proportion of the optical near-field information, and should therefore give the best spatial resolution, with the lowest background component. However, there is a signal-to-noise trade-off because the harmonic optical power decreases with n. Signals demodulated at the *n* = 2 harmonic were used in this study. To further decouple near-field from far-field, a pseudoheterodyne interferometric detection scheme was employed[28].

A full numerical analysis of the interaction between the laser radiation and the oscillating tip is beyond the scope of this paper. However, the fact that, in solid samples, damping broadens the IR absorption peaks means that typically the material dispersion is weak. This allows for an analytical model of the tip-sample interaction, (where the spherical tip is modelled as a point dipole) which, in turn, predicts that, the optical phase of the backscattered signal is proportional to the absorption coefficient of the material beneath the tip [15,19,20, 29].

Even if this "weakly dispersive" approximation is not reliable, given that the technique works with optical processes that are linear and therefore obey the principle of superposition, reliable concentration estimates can be obtained by measuring reference scattering phase values with the bare substrate, and subsequently recording the degree of phase shift as the tip is rastered over the bio-material [19,20].

To analyse the MICHNI images, the spectroscopic signature of the BTZ was isolated by first acquiring a frequency dependent reference phase value, $\varphi_{glass}(\nu)$, with the probe imaging a part of the glass substrate close to the ultrastructure component of interest. This allowed the degree of phase shift corresponding to the BTZ absorption to be computed as $(\varphi(\nu) = \varphi_{BTZ}(\nu) - \varphi_{glass}(\nu))$. The backscattered amplitude was then normalised to that of the glass, $(s(\nu) = s_{BTZ}(\nu)/s_{glass}(\nu))$. This allowed the imaginary component of the backscattered signal, $Im(\sigma)$ to be computed as $Im(\sigma) = s \sin(\varphi)$.

Here $Im(\sigma)$ is the imaginary component of the complex near-field scattering coefficient, which is itself proportional to the absorption coefficient of the material beneath the tip. For the BTZ experiments, the $Im(\sigma)$ signal showed the pronounced spectral resonance at 1020 cm$^{-1}$ shown in Fig.4 (d)).

Unfortunately, at the time of writing, this quantitative analysis cannot be performed in e.g. the depths of the cell nucleus because of difficulties in accounting for phase uncertainties caused by the large height variations across the nucleus.

**Far-field infrared absorption spectroscopy.**

A pure sample of BTZ was acquired from Selleckchem at 99.94% purity (catalogue number S1013, batch number S101315). A commercial FTIR spectrometer (Frontier FT-IR/MIR, PerkinElmer) equipped with a KBr window, a ATR (attenuated total reflectance) accessory and a LiTaO$_3$ detector was used to record the IR spectrum in the 4000-650 cm$^{-1}$ range, at 4 cm$^{-1}$ resolution and with a scan speed of 0.2 cm sec$^{-1}$). Approximately, 2 mg of solid BTZ was applied over the exposed KBr ATR crystal surface using a carefully cleaned sample mounting plate. The knob of the pressure arm was then swung over the sample and knob was rotated to apply pressure. Finally, the ATR spectrum of BTZ was recorded and normalised against a background air spectrum (see supporting information).

**Cell preparation.**

All the studies were performed using RPMI-8226 myeloma cells, cultured in RPMI-1640 medium. The medium was supplemented with 10% FCS, penicillin (100U/ml), 5% glutamine, and streptomycin sulphate (100µg/ml). Cells were cultured at 37°C in a humidified atmosphere containing 5% $CO_2$. Cells were harvested, washed twice with PBS, and counted prior to BTZ treatment. 1x10$^7$ cells were incubated with 8µM of BTZ in vehicle (0.1% DMSO) for 1h at 37°C. Cells were washed twice in PBS before being pelleted for paraffin embedding and sectioning.

Pellets of washed RPMI-8226 myeloma cells were fixed in 4% paraformaldehyde for 30min at room temperature. These were processed for embedding in paraffin as per published protocol[11]. Cell blocks were cut into ~1µm sections using a microtome before being mounted onto glass slides. To ensure the sections adhered well, the glass slides were heated to 60°C for 10min before 5min of cooling at room temperature. The mounted samples were dewaxed by being immersed in a series of two xylene, followed by three ethanol baths, each for 5min.

This sample preparation process is the formalin-fixed-paraffin-embedded (FFPE) "gold standard" protocol[11] that is used to make many tens of thousands of biopsy sections in hospital pathology laboratories for routine haematoxylin and eosin ("H+E") based disease diagnoses annually.

The principle modifications were to use a thinner section (estimated to be ~1μm thick before dewaxing,) so that the sample height fluctuations were kept within the allowable range of the AFM, and omitting the H+E staining and protection stages of the sample preparation protocol altogether.

The MICHNI images are all taken at ambient pressure and temperature, and the cells are unstained and unlabelled throughout.

At present the specimens are imaged while they are dry, so whilst the chemical information is well preserved, the details of the ultrastructural morphology will be influenced by shrinkage rates that may vary between the different ultrastructure components. This makes it difficult to be confident when comparing MICHNI images with those in the EM-based literature.

Looking ahead, we expect that a combination of operator experience, coupled with correlation studies and developments on fixing protocols will surmount this obstacle.

**Acknowledgements:-** CCP, WH and HA acknowledge financial support from EPSRC (EP/K503733/1, EP/K029398) MF acknowledges support from EPSRC (EP/L014580/1, EP/R00188X/1) and Cancer Research UK (C33325/A19435). Discussions with Jeremy Skepper are gratefully acknowledged. AR-Z acknowledges fellowship support from the Alfonso Martin Escudero Foundation. EA and AB acknowledge support from Cancer Research UK, C2536/A16584.

**Author Contributions:-** WH acquired and analysed the MICHNI images. JB, SS and AR-Z prepared and measured reference BTZ samples. AB, WH and HA supplied and prepared cell samples for MICHNI imaging. AP analysed the MICHNI images. MF selected and characterised the BTZ drug. CP conceived and managed the project. All authors contributed to the preparation of the manuscript.

**Competing Financial Interests.** None of the authors have competing financial interests.

# Supplementary information

**Comparison of Bortezomib and Human tissue infrared spectra**

Bortezomib (BTZ) is a promising candidate for spectroscopic mapping, as it has an unusual chemical structure, containing boron, and an associated B-C bond (Fig. S1(a)), which is rarely found in nature. This yields a strong, sharp absorption resonance in a spectral region where the absorption spectrum of biological material is normally featureless (Fig. S1(b)).

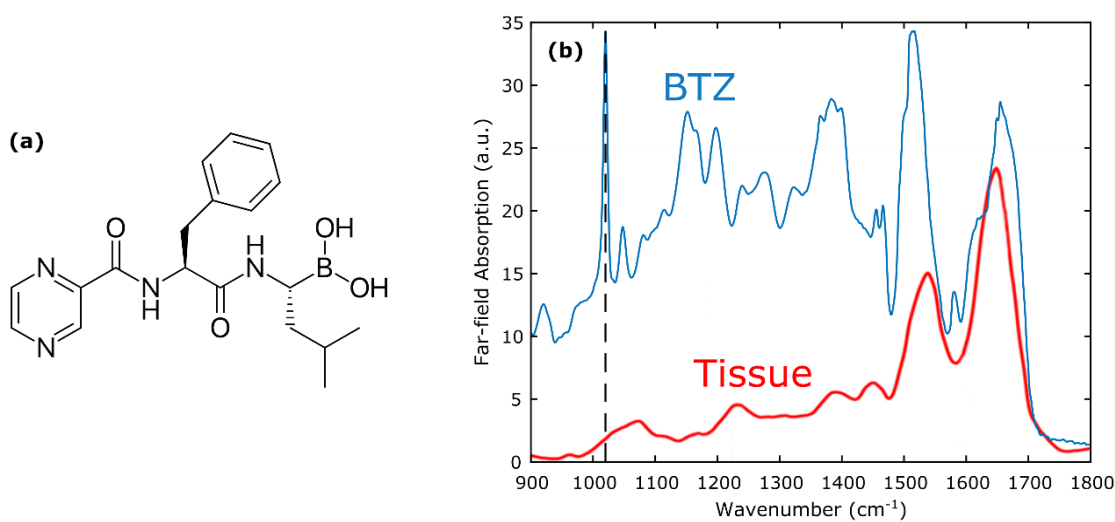

*Figure S1: (a) The unusual chemical structure of Bortezomib (BTZ) results in (b) an infrared far-field absorption spectrum (blue) that is highly distinctive compared with typical human tissue (red; reproduced with permission from [30]).*

**The relative contributions of the near-field amplitude and phase to spectral changes in Im(σ).**

Strictly speaking, it is the imaginary component of the near-field scattering that is proportional to local absorption[12], however it is often the case that the majority of the chemical contrast signal in the imaginary component spectrum comes from the contribution of the near-field phase variation. This is clearly shown in Fig. S2, whereby the amplitude spectrum is relatively flat compared with the phase spectrum. Comparing these figures with Fig. 4(d) of the main text, it is clear that the phase and imaginary spectra are in good agreement.

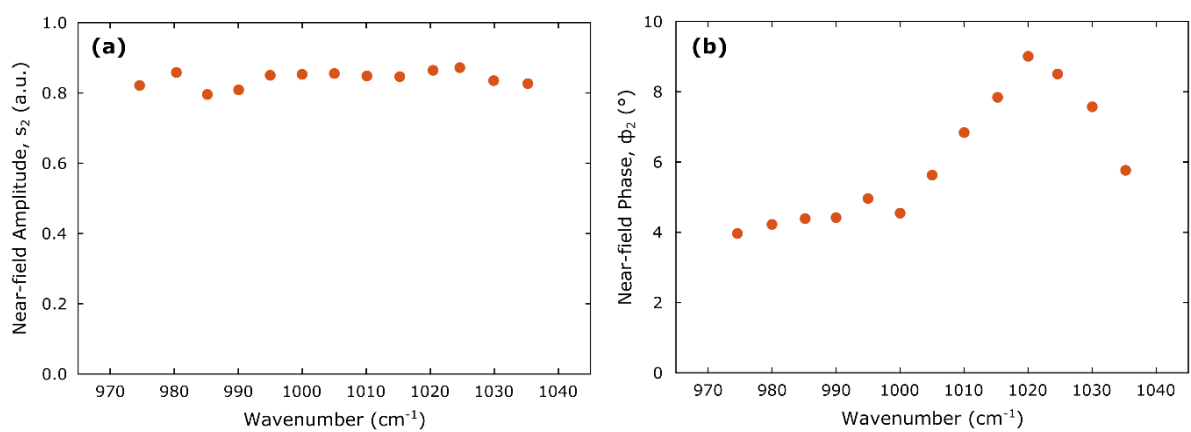

*Figure S2: (a) Near-field amplitude and (b) phase spectra of Bortezomib in a cell (as shown in Fig. 4.)*